\documentclass[reprint, superscriptaddress, amsmath,amssymb, aps, pra,]{revtex4-2}

\usepackage[utf8]{inputenc}
\usepackage{graphicx}
\usepackage{dcolumn}
\usepackage{bm}
\usepackage{hyperref}
\usepackage{braket}
\usepackage{booktabs,tabularx}

\usepackage{todonotes,xcolor}
\usepackage[normalem]{ulem}

\renewcommand{\d}{{\rm d}}
\newcommand{\im}{{\rm i}}

\newcommand{\tcut}{t_{\rm c}}
\newcommand{\teff}{\tilde{t}_0}

\newcommand{\D}[1]{{#1}^\dagger}
\newcommand{\fD}[1]{{#1}^{ \phantom{\dagger}}}

\newcommand{\bq}{{b {\bf q}}}
\newcommand{\wmin}{\bar{\omega}_{\rm low}}
\renewcommand{\IJ}{{IJ}}
\newcommand{\ab}{{\alpha \beta}}
\newcommand{\IJab}{_{\IJ}^{\ab}}

\usepackage{glossaries}
\setacronymstyle{long-short}
\newacronym{gk}{GK}{Green-Kubo}
\newacronym{aigk}{aiGK}{\emph{ab initio} Green-Kubo}
\newacronym{md}{MD}{molecular dynamics}
\newacronym{aimd}{aiMD}{\emph{ab initio} molecular dynamics}
\newacronym{pes}{PES}{potential-energy surface}
\newacronym{bte}{BTE}{Boltzmann transport equation}
\newacronym{pgm}{PGM}{phonon gas model}
\newacronym{mgo}{MgO}{magnesium oxide}
\newacronym{cui}{CuI}{copper iodide}
\newacronym{hfacf}{HFACF}{heat flux autocorrelation function}
\newacronym{gga}{GGA}{generalized gradient approximation}

\begin{document}

\title{Ab initio Green-Kubo simulations of heat transport in solids: Method and implementation}

\author{Florian Knoop}
\affiliation{
The NOMAD Laboratory at the FHI of the Max-Planck-Gesellschaft and IRIS-Adlershof of the Humboldt-Universität zu Berlin, 14195 Berlin, Germany}
\affiliation{Theoretical Physics Division, Department of Physics, Chemistry and Biology (IFM), Linköping University, SE-581 83 Linköping, Sweden}
\author{Matthias Scheffler}
\affiliation{
The NOMAD Laboratory at the FHI of the Max-Planck-Gesellschaft and IRIS-Adlershof of the Humboldt-Universität zu Berlin, 14195 Berlin, Germany}

\author{Christian Carbogno}
\affiliation{
The NOMAD Laboratory at the FHI of the Max-Planck-Gesellschaft and IRIS-Adlershof of the Humboldt-Universität zu Berlin, 14195 Berlin, Germany}

\begin{abstract}
  \emph{Ab initio} Green-Kubo~(aiGK) simulations of heat transport in solids allow for assessing lattice thermal conductivity in anharmonic or complex materials from first principles. 
In this work, we present a detailed account of their practical application and evaluation with an emphasis on noise reduction and finite-size corrections in semiconductors and insulators.
To account for such corrections, we propose strategies in which all necessary numerical parameters are chosen based on the dynamical properties displayed during molecular dynamics simulations in order to minimize manual intervention. This paves the way for applying the aiGK method in semi-automated and high-throughput frameworks. The proposed strategies are presented and demonstrated for computing the lattice thermal conductivity at room temperature in the mildly anharmonic periclase MgO, and for the strongly anharmonic marshite CuI. 
\end{abstract}

\date{September 2022}
\maketitle
\glsdisablehyper

\section{Introduction \label{sec:introduction}}
Heat transport is an important phenomenon in many branches of physics and adjacent fields, be it materials science investigating technologically relevant compounds~\cite{Snyder.2008,Perepezko.2009}, or astrophysics and earth sciences, where thermodynamic properties of planets are studied~\cite{Nimmo.2007,Grasselli.2020}.
In dielectric solids, thermal transport is mostly determined by the conduction of heat energy in the form of thermal nuclear motion~(lattice thermal conductivity), and electronic heat transport, photonic heat radiation, as well as convective contributions due to mass transport can be neglected~\cite{Vogelsang.1987}. 
\emph{Ab initio} simulations of the lattice thermal conductivity are typically performed in the framework of phonon theory: The \gls{pes} is approximated by force constants which can be obtained either as potential-energy derivatives or in a renormalized, temperature-dependent fashion~\cite{Esfarjani:2008iv,Hellman.2013,Paulatto.2015}. The equation of motion is solved for the harmonic, second-order terms, which results in decoupled phonon modes. Higher-order terms, up to fourth order~\cite{Feng.2017}, are included via perturbation theory to compute phonon lifetimes. The \gls{bte}~\cite{Broido.2007,Simoncelli.2019} then yields particle-like conduction contributions to the thermal conductivity. Additional contributions from wave-like conduction are accounted for in the Wigner transport formulation~\cite{Simoncelli.2022}. These contributions are particularly important in complex crystals when the individual phonon modes overlap. 

However, thermal insulators with $\kappa < 10$\,W/mK of importance for applications such as thermal barrier coatings in heat engines~\cite{Perepezko.2009,Clarke.2012} or thermoelectric materials for waste-heat recovery~\cite{Snyder.2008}, are often strongly anharmonic~\cite{Knoop.2020}, and the phonon picture underlying the Boltzmann or Wigner transport equations likely breaks down,  even when renormalized, temperature-dependent force constants are used~\cite{Xia.2020}. The Ioffe-Regel criterion~\cite{Ioffe.1960} poses a formal limit for the validity of the phonon picture and therefore perturbative formalisms~\cite{Simoncelli.2022}. 
This intuitive criterion states that phonons are only well-defined quasiparticles if their lifetimes exceed their oscillation periods. Non-perturbative approaches based on \gls{gk} theory~\cite{Green.1954,Kubo.1957,Kubo.1957oqc} do not suffer from this shortcoming since the heat flux is evaluated in molecular dynamics simulations and all anharmonic effects are taken into account. Accordingly, the \gls{gk} method covers the transport mechanisms described by the Boltzmann and Wigner transport equations, but also the regime beyond the Ioffe-Regel criterion in which the phonon picture becomes 
invalid~\cite{Simoncelli.2022,Caldarelli.2022bea}.
Its extension to first principles frameworks,~i.\,e.,~the \gls{aigk} technique, was introduced recently~\cite{Carbogno.2017,Marcolongo.2016}.
By evaluating the \emph{ab initio} heat flux along \gls{aimd} trajectories, \gls{gk} theory can be used to access the thermal conductivity in a non-perturbative way
on the basis of a fully \emph{ab initio} description of the \gls{pes}. Simplifying model assumptions about the \gls{pes} such as the \mbox{(quasi-)}harmonic approximation are therefore not needed. This makes \gls{aigk} a suitable tool for the parameter-free study of materials of arbitrary anharmonic strength whenever the nuclear dynamics 
can be described by \gls{aimd} simulations,~i.\,e.,~at temperatures where nuclear quantum effects can be 
neglected~\cite{Sutherland.2021}, and at which the system is not close to a structural phase transition~\cite{Chun.2021}.

While \gls{aigk} offers an encompassing framework for first-principles heat transport simulations, its practical implementation brings a set of challenges that need to be addressed:
The noise due to thermodynamic fluctuations in small ensemble sizes of less than ten \gls{aimd} trajectories with tens of picoseconds simulation time each, and finite-size effects when using supercells of hundreds atoms only.
These hurdles are of particular importance in crystalline systems where finite-size effects can be significant because readily accessible supercell sizes are not sufficient to describe all relevant length scales~\cite{Schelling.2002,Sellan.2010}, as opposed to amorphous systems and liquids where vibrations are more localized~\cite{Allen.1993}.

Several ideas to reduce noise~\cite{Marcolongo.2020,Ercole.2017} and account for finite-size effects~\cite{Carbogno.2017} have been suggested in the literature. Their application to practical calculations, however, requires educated choices for several numerical parameters. Furthermore, no unified framework addressing both problems is available to date, hindering broader application of aiGK methods for crystals.
In this work, we present and discuss such a framework and its numerical implementation in FHI-vibes~\cite{Knoop.2020cx} for two test systems: Periclase \gls{mgo}, and marshite \gls{cui}. Both are simple binary, cubic compounds, however, \gls{cui} is much more anharmonic~\cite{Knoop.2020}, and \gls{bte} simulations overestimate its thermal conductivity significantly~\cite{Togo.2015}.

We demonstrate the implementation in detail for the case of periclase \gls{mgo} which is well-known in the literature of first-principles heat transport techniques~\cite{Koker.2009,Stackhouse.2010,Tang.2010}.
We discuss the impact of noise-reduction and finite-size-extrapolation techniques, and propose strategies to apply such corrections without human intervention by choosing the required numerical parameters using the available \gls{aimd} data: First, we present a new real-time scheme to remove noise from the \gls{hfacf}, which enables to choose cutoff times in a numerically robust way based on a ``first dip'' criterion,~i.\,e.,~the time when the \gls{hfacf} drops to zero for the first time~\cite{Chen.2010}. The two-step procedure comprises discarding non-contributing terms from the flux based physical arguments~\cite{Ercole.2016}, and subsequent noise filtering that preserves the integrated thermal conductivity.
 Next, we discuss a size extrapolation scheme for periodic solids, adapted from the one first introduced in Ref.\,\cite{Carbogno.2017}, which allows to correct for finite size effects of simulation cells used in \gls{aimd} simulations. Finally, we discuss convergence in the simulation times. 
This approach is then applied to the strongly anharmonic \gls{cui}. Good agreement with the literature is obtained in both cases.

Both materials are studied at the level of the \gls{gga} using the PBEsol functional and light-default basis sets in FHI-aims~\cite{Perdew.2008,Blum.2009}. Supercells are $3 \times 3 \times 3$ extensions of the conventional, cubic unit cells, with 216 atoms each. The \gls{md} simulations are performed via FHI-vibes~\cite{Knoop.2020cx}. The aiGK method as described here is implemented in FHI-vibes as well. Force constants for the size extrapolation via harmonic mapping are obtained by regression from the \gls{md} runs via the temperature dependent effective potentials code (TDEP)~\cite{Hellman.2011,Hellman.2013}.
The \gls{md} runs are thermalized using the pre-thermalization technique outlined in Ref.\,\cite{West.2006} using finite-differences force constants obtained via phonopy~\cite{Togo.20154whg}. Afterwards, a Langevin thermostat at the target temperature (300\,K) is used to perform {\it NVT} sampling. After an initial sampling period of 2.5\,ps, the cell parameters are adjusted such that thermal pressure is minimized to below $5$\,kbar  in order to account for thermal expansion~\cite{Knoop.thesis}. Starting conditions for the {\it NVE} simulations are chosen from an {\it NVT} run for the relaxed supercell at least 2\,ps apart. The time step for the \gls{md} simulation was chosen as 5\,fs, which corresponds to a tenth of the shortest period duration of the harmonic spectrum of \gls{mgo} ($\omega_{\rm max} \approx 20$\,THz). The heat flux is sampled less frequently since heat transport is dominated by the slow vibrations, and all results are reported for a heat-flux sampling period of 20\,fs. We have checked that further decreasing the heat-flux sampling frequency does not change the results significantly.

The work is organized as follows: In Sec.\,\ref{sec:gk_theory}, we review Green-Kubo theory in order to highlight the steps necessary for the numerical implementation. Section\,\ref{sec:cutoff.time} presents our approach for noise reduction based on physical arguments and real-time signal analysis which allows to truncate the \gls{gk} time integral in a numerically robust way. Section~\ref{sec:size_extrapolation} presents the updated version of the size-extrapolation scheme first introduced in Ref.\,\cite{Carbogno.2017}. To complete the method description, we discuss results for \gls{mgo} in Sec.\,\ref{sec:time_convergence}, and compare to available experimental and computational literature. After completing the discussion of the method for \gls{mgo}, we apply the scheme to \gls{cui} in Sec.\,\ref{sec:cui}, and conclude with some remarks on simulation time convergence in Sec.\,\ref{sec:simulation.times}.

\section{Green-Kubo theory \label{sec:gk_theory}}
Let us start with a short summary of classical thermal transport in the framework of \gls{gk} theory~\cite{Kubo.1957,Kubo.1957oqc,Baroni.2018}:
The thermal conductivity tensor at temperature $T$ is given as the canonical ensemble average defined by the phase-space integral
\begin{align}
	\kappa^{\alpha \beta} (T)
		= \frac{1}{\mathcal Z} 
		\int \d \Gamma ~ \kappa^{\alpha \beta} [\Gamma] 
		~{\rm e}^{- \frac{1}{k_{\rm B} T} \mathcal H [\Gamma]}~,
	\label{eq:kappa.GK}
\end{align}
where \mbox{$\Gamma \equiv ({\bf R}_1, \ldots {\bf R}_N; {\bf P}_1, \ldots {\bf P}_N)$} are phase-space configurations for $N$ atoms with positions ${\bf R}_I$ and momenta ${\bf P}_I$. $\mathcal H [\Gamma]$ is the Hamiltonian of the system with corresponding partition function $\mathcal Z$, $k_{\rm B}$ is the Boltzmann constant, and $\alpha, \beta$ denote the Cartesian components of the tensor.
For each phase-space configuration $\Gamma$, the thermal conductivity is computed as
\begin{align}
	\kappa^{\alpha \beta} [\Gamma]
		&=
		\frac{V}{k_{\rm B} T^2} 
		\lim_{t_{\rm c} \to \infty}
		\int_{0}^{\tcut} 
		\d t ~ C_{JJ}^{\alpha \beta} [\Gamma] (t)~,
	\label{eq:kappa.trunc.t}
\end{align}
with the \gls{hfacf},
\begin{align}
\begin{split}
  	C^{\alpha \beta}_{J J} [\Gamma](t)
	&=
		\lim_{t_{0} \to \infty}
		\frac{1}{t_0 - t}
		\int_{0}^{t_{\rm 0} - t} 
		\d s ~ J^\alpha [\Gamma ({t + s})] J^\beta [\Gamma (s)] \\
	&\equiv \langle J^\alpha (t) J^\beta (0) \rangle~,
\end{split}
	\label{eq:hfacf}
\end{align}
where the phase-space points $\Gamma (t)$ in the trajectory are obtained from the time evolution generated by the many-body Hamiltonian of the system, $\mathcal H [\Gamma]$, by propagating the initial configuration $\Gamma (0)  \equiv \Gamma$ for a time $t$.
$J^\alpha [\Gamma(t)] \equiv J^\alpha (t)$ is the the heat flux component evaluated for the configuration $\Gamma (t)$,
and $\langle \cdot \rangle$ is the shorthand notation for the time average in Eq.\,\eqref{eq:hfacf}.

In order to evaluate these equations in finite simulations, the integrals need to be discretized and truncated to finite domains. First, Eq.\,\eqref{eq:kappa.GK} is approximated by taking a finite set of $M$ starting configurations $\Gamma^i (0) \equiv \Gamma^i$, so that
\begin{align}
	\kappa^{\alpha \beta} (T)
		\approx
		\frac{1}{M} \sum_{i=1}^M \kappa^{\alpha \beta} [\Gamma^i]~,
	\label{eq:kappa.mean}
\end{align}
where the starting conditions $\Gamma^i$ are chosen from {\it NVT} \gls{md} simulations for the thermodynamic conditions of interest. For each starting condition $\Gamma^i$, {\it NVE} \gls{md} simulations are performed to generate the time evolution of the system, $\Gamma^i (t)$, and evaluate the heat flux ${\bf J} (t)$ along this trajectory. The simulation is performed for a total simulation time $t_0$, thereby truncating the time integral in Eq.\,\eqref{eq:hfacf}. This time needs to be large enough to cover the time scales of the physical processes relevant for heat transport.
 From the resulting autocorrelation function of finite length, the thermal conductivity components are computed via Eq.\,\eqref{eq:kappa.trunc.t}. For each component, a \emph{cutoff time} $\tcut < t_0$ is chosen to avoid integrating parts of the \gls{hfacf} after it has effectively decayed, since its tail can be heavily affected by statistical fluctuations stemming from finite size and time effects~\cite{Zwanzig.1969,Jones.2012,Oliveira.2017,Wang.2017wl}, or tiny but systematic drifts that accumulate in the long time limit,~e.\,g.,~when the average flux $\langle {\bf J} \rangle_t$ does not vanish exactly over the simulation time or appears slightly skewed.

After computing the thermal conductivity tensor for each trajectory, the final value is given by Eq.\,\eqref{eq:kappa.mean},~i.\,e.,~by the \emph{mean} of the individual trajectories. The statistical error due to the finite ensemble average is estimated by the \emph{standard error},~i.\,e.,~the standard deviation of the mean,
\begin{align}
	\Delta \kappa^{\alpha \beta} (T)
		= \frac{1}{\sqrt{N}} \sqrt{\frac{1}{N} \sum_i \left( \kappa^{\alpha \beta} (T) - \kappa^{\alpha \beta} [\Gamma^i] \right)^2}~.
	\label{eq:kappa.error}
\end{align}
From the Cartesian components of the thermal conductivity $\kappa^{\alpha \beta} (T)$, the scalar thermal conductivity $\kappa (T)$ is obtained via
\begin{align}
	\kappa (T)
		= \frac{1}{3} \sum_{\alpha} \kappa^{\alpha \alpha} (T)~.
	\label{eq:kappa.scalar}
\end{align}
In principle, these equations can be evaluated as is, and convergence in size and time can be checked by simply increasing the respective scales. 
While this is computationally possible when using analytical force fields~\cite{He.2012}, this is certainly not desirable in the
\emph{ab initio} case, where the cost per time step is considerably higher:
Here, the accessible size and time scales are typically orders of magnitude lower, and additional steps to increase the amount of information that can be extracted from the comparatively short simulations are pivotal.

\section{Cutoff time and noise reduction \label{sec:cutoff.time}}
For a robust identification of the cutoff time $\tcut$, we first reduce noise from the \gls{hfacf} as much as possible.  This is achieved in two steps: First, we re-define the \emph{ab initio} heat flux used in this work such that terms not contributing to the thermal conductivity are discarded~\cite{Ercole.2016}. Second, we filter remaining contributions from the \gls{hfacf} that do not contribute to the integrated thermal conductivity. This allows to determine the cutoff time even in the presence of noise based on a ``first dip'' criterion,~i.\,e.,~by choosing the time when the \gls{hfacf} drops below zero for the first time.

The raw \emph{ab initio} virial heat flux used in this work was introduced in Ref.\,\cite{Carbogno.2017} and is given for a phase-space point $\Gamma (t) = \set{{\bf R} (t), {\bf P} (t)}$ by
\begin{align}
  {\bf J}_{\rm raw} [\Gamma (t)]
    \equiv {\bf J}_{\rm raw} (t) 
		= \frac{1}{V} \sum_I \sigma_I (t)  \dot{\bf R}_I (t)~,
	\label{eq:J.0}
\end{align}
where $\sigma_I (t) \equiv \sigma_I [{\bf R} (t)]$ is the contribution of atom $I$ to the virial stress tensor for the configuration ${\bf R} = ({\bf R}_1, \ldots {\bf R}_N )$ at the given time $t$ as derived and discussed in Ref.\,\cite{Carbogno.2017,Knuth.2015}, and $\dot{\bf R}_I (t) = {\bf P}_I (t) / M_I$ is the velocity of atom $I$ with mass $M_I$. In this definition of the heat flux, convective contributions that become important in liquids and gases are entirely neglected~\cite{Vogelsang.1987,Ercole.thesis}.

\subsection{Discarding non-contributing terms}
\begin{figure}
	\includegraphics[width=\columnwidth]{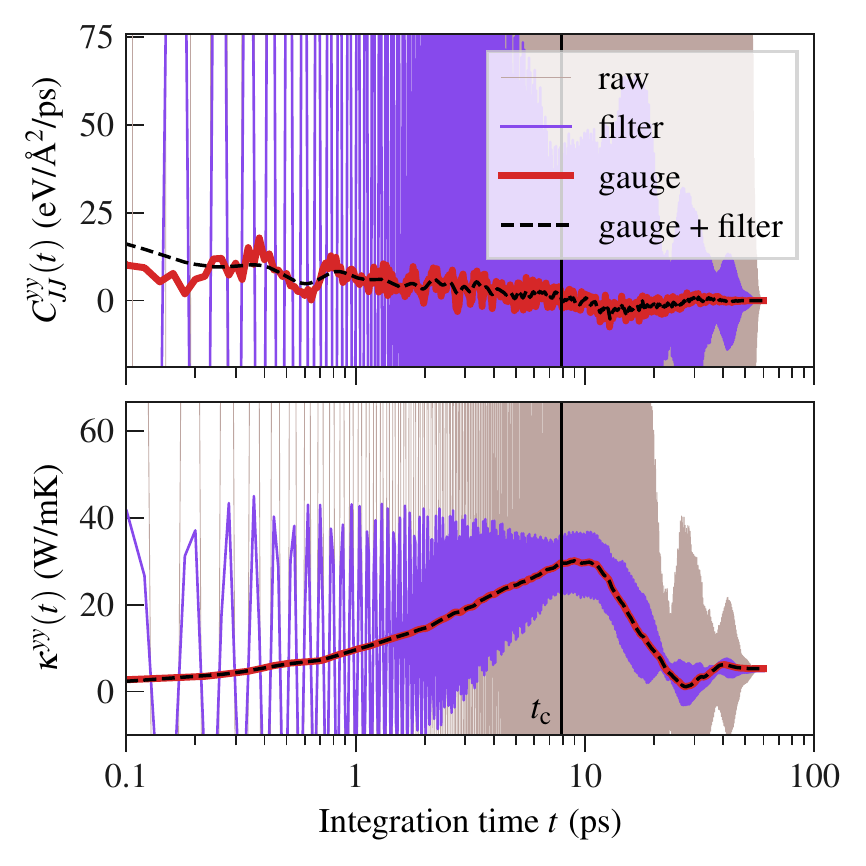}
	\caption{Heat flux autocorrelation function (HFACF) $C^{yy}_{JJ}(t)$ in \gls{mgo} for the $yy$-component as defined in Eq.\,\eqref{eq:hfacf}, and its cumulative integral,~i.\,e.,~the thermal conductivity $\kappa^{yy} (t)$ defined in Eq.\,\eqref{eq:kappa.cum} as function of integration time $t$. Shaded gray: $C^{yy}_{JJ}(t)$ and $\kappa^{yy} (t)$ obtained by using the raw flux as defined in Eq.\,\eqref{eq:J.0}. Purple: Using only the filter discussed in Sec.\,\ref{sec:noise.filtering}. Red: Using the gauge-fixed flux defined in Eq.\,\eqref{eq:J.1}. Black dashed curves: After applying noise filtering to the gauge-fixed quantities as explained in the main text. The cutoff time $t_{\rm c}$ is chosen based on the ``first dip'' of the gauge-fixed and noise-filtered HFCAF.
	}
	\label{fig:imp.hfacf.kappa.1}
\end{figure}

We split the raw flux in two parts,
\begin{align}
	{\bf J}_{\rm raw} (t)
		= \underset{1)}{\underbrace{\frac{1}{V} \sum_I \delta \sigma_I (t) \dot{\bf R}_I (t) }}
		+ \underset{2)}{\underbrace{\frac{1}{V} \sum_I \braket{\sigma_I}_t \dot{\bf R}_I (t)}}~,
	\label{eq:J.1}
\end{align}
where $\braket{\sigma_I}_t$ is the time-averaged atomic virial, and $\delta \sigma_I (t)$ is the time-dependent part. In the absence of diffusion, the second term is the total time derivative of a bounded vector field, $\sum_I \braket{\sigma_I} \dot{\bf R}_I (t) = \frac{\d}{\d t} \sum_I \braket{\sigma_I} {\bf U}_I (t)$, where ${\bf U}_I (t) = {\bf R}_I (t) - {\bf R}_I$ is the displacement of atom $I$ from its reference position in the lattice, ${\bf R}_I$. Contributions to the heat flux that can be written in this form do not contribute to the integrated transport coefficient, as can be elegantly shown using the ``gauge theorem'' discussed in detail in Ref.\,\cite{Ercole.2016,Isaeva.2019,Marcolongo.2020}. We therefore discard the second term from the flux, and proceed using the following gauge-fixed heat flux expression:
\begin{align}
	{\bf J} (t)
		= \frac{1}{V} \sum_I \delta \sigma_I (t) \dot{\bf R}_I (t)~.
	\label{eq:J}
\end{align}
In compounds with two or more elements, the individual average virials $\langle \sigma_I \rangle$ can be significant, and discarding the non-contributing part from the raw heat flux reduces the noise in the \gls{hfacf} considerably, as shown for the case of \gls{mgo} in Fig.\,\ref{fig:imp.hfacf.kappa.1} (red curves compared to gray curves). We note that this amount of noise reduction is difficult to achieve by means of mere filtering: The blue curves in Fig.\,\ref{fig:imp.hfacf.kappa.1} are obtained by filtering the raw \gls{hfacf} obtained with the flux defined in Eq.\,\eqref{eq:J.1} with the filter discussed below in Sec.\,\ref{sec:noise.filtering}. It is apparent that leveraging the gauge theorem by using the flux defined in Eq.\,\eqref{eq:J} instead reduces the variance in the \gls{hfacf} much more significantly, and is furthermore physically rigorous.
Finally, we enforce a vanishing expectation of the flux to remove bias from the resulting quantities due to the finite time of the simulation by removing the finite-time average, ${\bf J} (t) \to \delta {\bf J} (t) = {\bf J} (t) - \braket{{\bf J}}_t$.

We note in passing that the above argument holds for any heat flux that can be written in a virial-based form similar to Eq.\,\eqref{eq:J.0},~e.\,g.,~common heat flux expressions for empirical or machine-learned force fields~\cite{Fan.2015}. For such force fields, a noise-reduction approach similar to the one presented here was very recently developed independently and applied successfully by Pereverzev and Sewell~\cite{Pereverzev.2022}.

\subsection{Noise filtering \label{sec:noise.filtering}}
After obtaining the gauge-fixed heat flux by discarding the non-contributing term, there is still a considerable level of noise in the HFACF which hinders a robust identification of the time at which it is effectively decayed,~i.\,e.,~the cutoff time $\tcut$. Available techniques to identify cutoff times, such as the first avalanche method introduced in Ref.\,\cite{Chen.2010} typically require system-dependent parameters, such as a tolerable signal-over-noise ratio or window sizes for moving average computation. To overcome this issue, we suggest an approach that does rely only on one single parameter which is chosen based on the vibrational spectrum of the material: Motivated by the fact that the \emph{integrated} HFACF,~i.\,e.,~the cumulative thermal conductivity
\begin{align}
	\kappa (t)
		=
		\frac{V}{k_{\rm B} T^2} 
		\int_{0}^{t} 
		\d t' ~ C_{JJ} (t')~,
	\label{eq:kappa.cum}
\end{align}
is already a much smoother function than the HFACF itself, we apply a moving window average to $\kappa (t)$ instead of $C_{JJ} (t)$.
The remaining parameter is the window size for the filter. It is chosen based on the vibrational spectrum of the material by taking the period length corresponding to the slowest significant frequency, $t_{\rm window} = 1/ \omega_{\min}$, which is chosen to be the first peak in the vibrational density of states (VDOS).
To ensure that $\kappa (t)$ vanishes identically at $t=0$, the filter is applied to the cumulative thermal conductivity extended antisymmetrically to negative times via $\kappa (-t) = - \kappa (t)$, a property which follows from the time symmetry of $C_{JJ} (t)$~\cite{Coretti.2018}.
Thereby all noise and non-contributing parts of higher frequency are effectively filtered from $\kappa (t)$, while all relevant time integrals are preserved by construction.
As required, the cumulative kappa before (red curve) and after filtering (black dashed curve) lie right on top of each other in the lower panel of Fig.\,\ref{fig:imp.hfacf.kappa.1}.
As in the previous section, we stress that using the gauge-fixed heat flux is crucial to access the integrated $\kappa (t)$, as shown in the lower panel of Fig.\,\ref{fig:imp.hfacf.kappa.1} in comparison to the case where only the filter was used to smoothen $\kappa (t)$ obtained from the raw flux (blue curve).

The filtering is carried over to the \gls{hfacf}, $C_{JJ} (t)$, by numerically differentiating the filtered cumulative thermal conductivity with respect to time and applying the same filter on the numerical gradient of $\kappa (t)$. The resulting \gls{hfacf} $C_{JJ} (t)$ is shown as a black dashed curve in the upper panel of Fig.\,\ref{fig:imp.hfacf.kappa.1}, and the further reduced level of variance in the \gls{hfacf} is apparent.
As seen in Fig.~\,\ref{fig:imp.hfacf.kappa.1},  $\kappa (t)$ reaches a plateau at approximately 8\,ps that lasts for several picoseconds. After that, numerical noise dominates and the accumulated numerical errors lead to a drop in  $\kappa (t)$. The ``first dip'' criterion is used to detect this plateau numerically in a reliable fashion. For this purpose, a cutoff time $\tcut$ is chosen that corresponds to the time when the signal-over-noise ratio vanishes, i.\,e., when  $C_{JJ} (t)$ drops to zero~\cite{Chen.2010}. Note that also different numerical approaches~\cite{Chen:2010kc,Oliveira.2016,Ercole.2017,Jones:2012dx} or a visual inspection of  $\kappa (t)$ can be used to identify this plateau. The main advantage of the ``first-dip'' criterion used here is that no numerical parameters have to be chosen, which facilitates automatic evaluation and the systematic comparison of aiGK simulations with different trajectory lengths~$t_0$. In turn, this results in smoothly converging thermal conductivities with respect to~$t_0$ as shown below in see Fig.~\ref{fig:kappa_convergence.mgo} and~\ref{fig:kappa.convergence.CuI}. This reflects that the final, converged values for $\kappa (t)$ are virtually independent from the details used to choose the cutoff time~$\tcut$.
With the cutoff time $\tcut$, the resulting thermal conductivity for a given component of the thermal conductivity tensor is given by the value $\kappa = \kappa (\tcut)$ as indicated by the horizontal line in Fig.\,\ref{fig:imp.hfacf.kappa.1}.
The presented scheme will be used for all reported values of thermal conductivity in the following.

We note that this filter corresponds to a low-pass filter in Fourier space~\cite{Ercole.2017}. However, since we found the real-time noise-reduction scheme as presented above sufficient to obtain robust results, we did not investigate additional processing steps that involve discrete (inverse) Fourier transforms with further numerical parameters and potential aliasing problems when only a few thousand data points are available.

\section{Size extrapolation for crystals\label{sec:size_extrapolation}}
After we have seen how the \gls{gk} formula is used to compute thermal conductivities from the \emph{ab initio} heat flux evaluated along \gls{aimd} trajectories, we discuss an update to the size-correction scheme for crystals first introduced in Ref.\,\cite{Carbogno.2017}.
The aim of this extrapolation is to correct for size effects occurring in \gls{aimd} simulations because phonon modes of longer wavelength than the supercell dimensions are not included.
This is of particular importance in crystals with periodic long-range order where these modes can contribute significantly to heat transport. 
The correction works by computing the harmonic contribution to the thermal conductivity $\kappa_{\rm ha}$ within the supercell via 
\begin{align}
	\kappa_{\rm ha}^{\alpha \beta} = V k_{\rm B} \sum_\bq v^\alpha_\bq v^{\beta}_\bq \fD \tau_\bq~,
	\label{eq:K.bte}
\end{align}
where $V$ is the system volume, $k_{\rm B}$ is the classical heat capacity per phonon mode, $v_\bq$ is the group velocity of a phonon mode with band index $b$ and wave vector $\bf q$, and $\tau_\bq$ is the lifetime of the mode extracted from the \gls{aimd} trajectory. This contribution is first computed for the wave vectors $\bf q$ commensurate with the supercell, $\kappa_{\rm ha - supercell}$, and then \emph{extrapolated} to bulk limit by \emph{interpolating} the lifetimes to denser $\bf q$-meshes in the Brillouin zone, $\kappa_{\rm ha - bulk}$. The resulting size-corrected thermal conductivity is obtained as
\begin{align*}
  \kappa_{\rm corrected}
    = \kappa + \underset{\delta \kappa_{\rm ha-correction}}{\underbrace{\kappa_{\rm ha - bulk} - \kappa_{\rm ha}}}~.
\end{align*}
The necessary steps to compute $\kappa_{\rm ha}$, $\kappa_{\rm ha - bulk}$, and therefore $\delta \kappa_{\rm ha-correction}$ are presented below.

In passing, we like to mention the main differences to the original approach~\cite{Carbogno.2017}. Space-group symmetries are now systematically exploited and all reciprocal-space quantities are generated from the irreducible part of the Brillouin zone. This guarantees unambiguous branch matching during the interpolation. Further minor changes are that lifetimes are extracted from fitting an exponential decay, cf.~Sec.\,\ref{sec:extrapolation.lifetimes}, and scaled lifetimes are interpolated linearly instead of using Fourier interpolation, cf.~Sec.\,\ref{sec:extrapolation.interpolation}.

\subsection{Harmonic mapping}
In order to map the real-space dynamics to the phonon picture which allows for interpolating in reciprocal space, we first define an auxiliary harmonic model determined by the real-space dynamical matrix
\begin{align}
  D\IJab
    = \frac{1}{\sqrt{M_I M_J}} \Phi\IJab~,
  \label{eq:D_IJ}
\end{align}
where $\Phi\IJab$ are the $(\alpha, \beta)$ components of harmonic force constants in the supercell between atom pairs $(I, J)$. As noted earlier, we obtain the force constants via the TDEP method to account for finite-temperature renormalization of phonon frequencies, eigenvectors, and group velocities~\cite{Hellman.2011,Hellman.2013}.

Using the crystal periodicity, the Fourier-transformed dynamical matrix reads
\begin{align}
  D_{{\bf q}, ij}^\ab
    = \sum_L {\rm e}^{\im {\bf q} \cdot ({\bf R}_i - {\bf R}_j - {\bf R}_L )}
    D_{i0, jL}^\ab ~,
  \label{eq:D_q}
\end{align}
where $\set{ {\bf R}_i, {\bf R}_j }$ denote reference positions in the unit cell, ${\bf R}_L$ is a Bravais lattice vector, and $\bf q$ is a commensurate wave vector fulfilling ${\bf q} \cdot {\bf R}_L = 2 \pi n$ with an integer $n$. 
The dynamical matrix yields real eigenvalues $\omega^2_\bq$ and complex eigenvectors ${\bf e}_{\bq, i}$ via the eigenvalue equation
\begin{align}
  \sum_{j \beta} D_{{\bf q}, ij}^\ab ~ e_{\bq, j}^\beta
    = \omega^2_\bq e_{\bq, i}^\alpha~,
  \label{eq:D_q.eval-equations}
\end{align}
where the band index $b$ was introduced to discern branches of solutions. To directly translate between real space coordinates $I = (i, L)$ and reciprocal space coordinates $b, {\bf q}$, we define the generalized eigenvector
\begin{align}
  {\bf e}_{\bq, I}
    \equiv \frac{1}{\sqrt{N_{\bf q}}} {\rm e}^{-\im {\bf q} \cdot {\bf R}_I}\, {\bf e}_{\bq, i}
  \label{eq:e_bqIa}
\end{align}
with ${\bf R}_I = {\bf R}_i + {\bf R}_L$, which diagonalizes the real-space dynamical matrix $D_{IJ}$ defined in Eq.\,\eqref{eq:D_IJ}, where $N_{\bf q}$ is the number of lattice points or commensurate wave vectors in the supercell, respectively.

Using the generalized eigenvector defined in Eq.\,\eqref{eq:e_bqIa}, we define normal coordinates $u_\bq$ and  $p_\bq$ as
\begin{align}
\begin{split}
u_\bq (t)
  &= \sum_{I} \sqrt{M_I} \, {\bf e}_{\bq, I} \cdot {\bf U}_{I} (t) ~,\\
p_\bq (t)
  &=\sum_{I} \frac{1}{\sqrt{M_I}} \,{\bf e}_{\bq, I} \cdot {\bf P}_{I} (t)~,
\label{eq:p_st.periodic}
\end{split}
\end{align}
where ${\bf U}_I (t) = {\bf R}_I (t) - {\bf R}_I$ is the instantaneous displacement of atom $I$ from its reference position ${\bf R}_I$, and ${\bf P}_I (t)$ is its momentum as before.
From here, the time-dependent complex mode amplitude $a_\bq (t)$ follows~\cite{Born.1954},
\begin{align}
	a_\bq (t)
	&= \frac{1}{\sqrt{2}} \left( u_\bq (t) + \frac{\im}{\omega_\bq} p_\bq (t) \right)~,
	\label{eq:a_st}
\end{align}
from which the time-dependent mode-resolved energy is obtained,
\begin{align}
  E_\bq (t) = \omega_\bq^2 \, \D a_\bq (t) \fD a_\bq (t)~,
  \label{eq:E_bq(t)}
\end{align}
where we note in passing that the harmonic energy expression familiar from quantum mechanics is recovered by substituting $a_\bq \to  \sqrt{ \hbar / \omega_\bq} \, a_\bq$.

Using the mode-resolved energy, the harmonic heat flux can be defined as~\cite{Hardy.1963}
\begin{align}
  {\bf J}_{\rm ha} (t)
    = \frac{1}{V} \sum_\bq E_\bq (t) {\bf v}_\bq ~.
  \label{eq:J.ha}
\end{align}
For a classical harmonic system, we can use that $\langle E^2_\bq \rangle = (k_{\rm B} T)^2$ and that cross correlations between different modes $(b, {\bf q}) \neq (b', {\bf q}')$ vanish, so that the harmonic thermal conductivity follows by application of Eq.\,(\ref{eq:kappa.trunc.t})-(\ref{eq:hfacf}),
\begin{align}
  \kappa^\ab_{\rm ha}
    = V k_{\rm B} \sum_\bq v^\alpha_\bq v^{\beta}_\bq \int_0^\infty G_\bq (t)~,
  \label{eq:kappa.ha.1}
\end{align}
with the mode-energy autocorrelation function
\begin{align}
	G_\bq (t)
		= \frac{\braket{E_\bq (t) E_\bq (0)}}{\braket{E_\bq^2}}~.
	\label{eq:G_b}
\end{align}
The lifetime $\tau_\bq$ in Eq.\,\eqref{eq:K.bte} is therefore defined by the integral
\begin{align}
  \tau_\bq
    \equiv \int_0^\infty G_\bq (t)~.
  \label{eq:tau.bq}
\end{align}

\subsection{Lifetime extraction \label{sec:extrapolation.lifetimes}}
For a purely harmonic system, the mode-energy autocorrelation function $G_\bq (t)$ does not decay, and the lifetime given by Eq.\,\eqref{eq:tau.bq} diverges. In the presence of phonon-phonon interactions due to anharmonicity however, $G_\bq (t)$ decays and the resulting lifetime is finite.

\begin{figure}
	\includegraphics[width=\columnwidth]{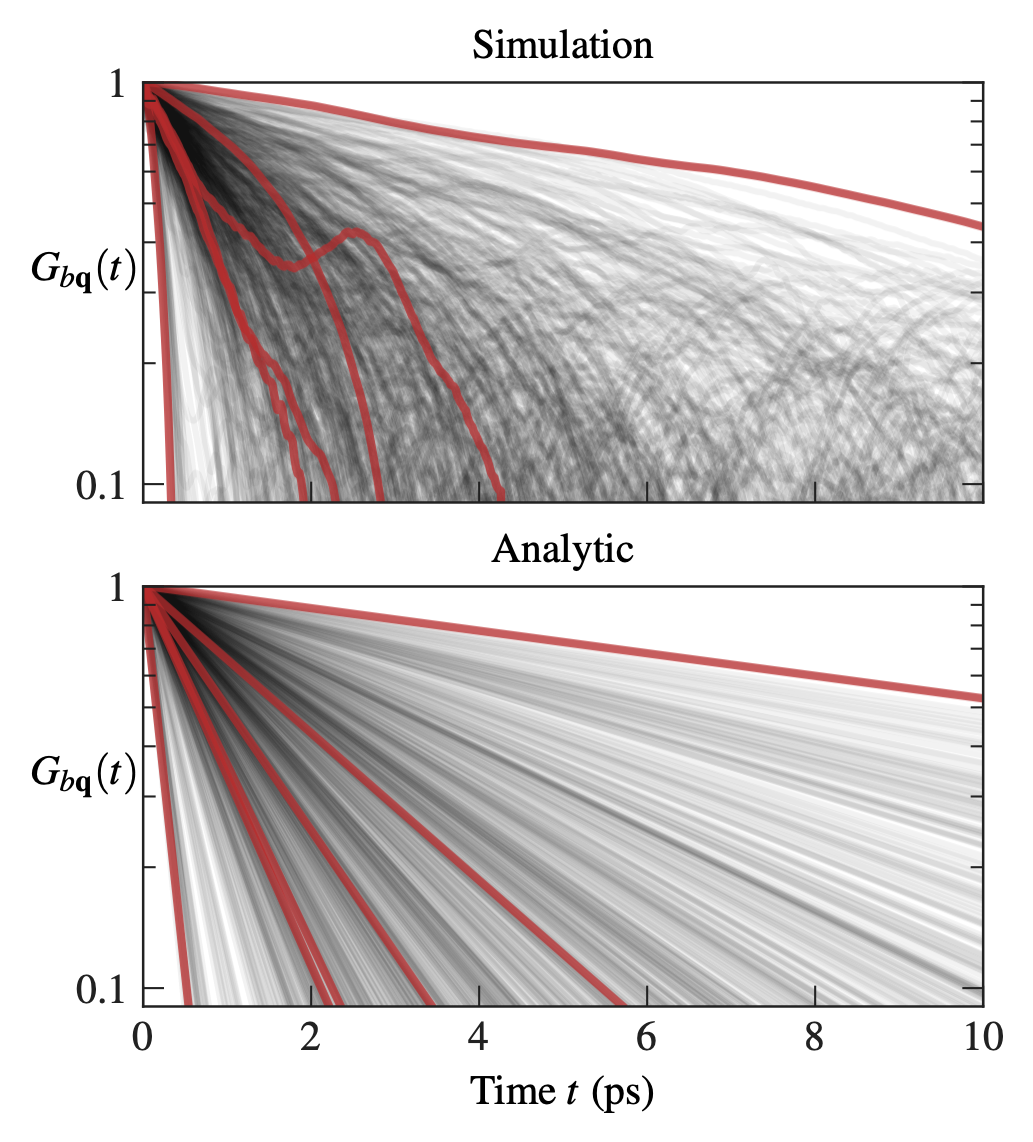}
	\caption{Fit of mode lifetimes for \gls{mgo} at 300\,K. Simulation performed with a timestep of 5\,fs for a simulation time of 60\,ps. Top: Normalized mode-energy autocorrelation function $G_\bq (t)$ as obtained from the simulation by Eq.\,\eqref{eq:G_b}. Bottom: Analytic expression given by Eq.\,\eqref{eq:G_b.approx} after fitting mode lifetimes $\tau_\bq$. The correlation functions for the six modes at ${\bf q} = (-1/6, -1/6, 0)$ are highlighted in red for comparison for times $t < 3 \tau_\bq$. The y-axis is logarithmic such that exponential functions appear as straight lines.
	}
	\label{fig:G_s}
\end{figure}

In order to evaluate the lifetime $\tau_\bq$ via Eq.\,\eqref{eq:tau.bq}, it is again necessary to integrate an autocorrelation function. As already mentioned in the introduction and Sec.~\ref{sec:cutoff.time}, this typically requires to choose an appropriate cutoff time to obtain numerically stable results. 
In the case of the lifetimes $\tau_\bq$, we exploit the advantage that we can integrate distinct phonon modes individually, in contrast to Sec.\ref{sec:cutoff.time} in which the flux for the whole system was processed at once.
In perturbation theory, the leading contribution to the decay of each phonon mode can be approximated via~\cite{Negele.1988}
\begin{align}
	G_\bq (t)
	 \approx {\rm e}^{- 2 {\rm Im} \Sigma_\bq t}
	\equiv {\rm e}^{-t / \tau_\bq} ~,
	\label{eq:G_b.approx}
\end{align}
where ${\rm Im} \Sigma_\bq$ is the imaginary part of the phonon self energy, and $\tau_\bq = 1 / 2 {\rm Im} \Sigma_\bq$ is the corresponding lifetime.
Under this approximation, the integral can be performed analytically based on the early decay of $G_\bq (t)$, hence allowing to capture also those long-lived, long-wavelength modes that are not guaranteed to be accessible via brute-force integration on the simulation time scales accessible in \gls{aimd} simulations. 
We compute $G_\bq (t)$ for each mode in the supercell, and obtain the corresponding lifetime by fitting Eq.\,\eqref{eq:G_b.approx} for times where $G_\bq (t) > 0.1$, in order to avoid fitting noise when $G_\bq (t)$ is effectively decayed. 
A comparison of numeric correlation functions via Eq.\,\eqref{eq:G_b}, and the respective analytic correlation functions given by Eq.\,\eqref{eq:G_b.approx}, is shown in Fig.\,\ref{fig:G_s}. We find that the analytic behavior,~i.\,e.,~exponential decay of $G_\bq (t)$, is indeed observed for many modes, in particular those with long lifetime. Some modes deviate more strongly from the exponential decay, for example the mode highlighted in Fig.\,\ref{fig:G_s} where $G_\bq (t)$ shortly increases after about 2\,ps, before dropping off again. The integrated correlation function,~i.\,e.,~the lifetime, is however only mildly affected from wiggles like this.

\subsection{Lifetime interpolation \label{sec:extrapolation.interpolation}}
For a given simulation $\set{\Gamma^i (t)}$, the lifetimes $\tau_\bq$ are evaluated for all commensurate $\bf q$-points, and projected to the symmetry-inequivalent points in the Brillouin zone determined by the space group operations of the system to improve the statistics: The irreducible {\bf q}-points in the Brillouin zone are obtained by iteratively reducing the given grid with the available symmetry operations for the system obtained by the spglib package~\cite{Togo.2018}. To avoid band-index matching problems between different $\bf q$-points, the eigenvectors for the full grid of commensurate $\bf q$-points are created by solving the eigenvalue problem in Eq.\,\eqref{eq:D_q.eval-equations} on the irreducible grid, and transforming the eigenvectors to the reducible points according to the transformation rules given in Ref.\,\cite{Maradudin.1968}.

In the next step, the lifetimes $\tau_\bq$ are interpolated to denser $\bf q$-point meshes. For this purpose, the fully anharmonic lifetimes $\tau_\bq$ at the commensurate $\bf q$-points are used to define one function $\lambda_b ( {\bf q})$ for each branch $b$ such that 
\begin{align}
    \fD{{\tau}}_b ({\bf q}) = \fD \lambda_b ( {\bf q}) \omega_b^{-2} ( {\bf q})~.
    \label{eq:tau_lambda}
\end{align}
The frequency scaling ensures that $\lambda_b ( {\bf q})$ is only weakly ${\bf q}$-dependent, which facilitates a linear interpolation of the lifetimes to arbitrary values $\tilde {\bf q}$ in the Brillouin zone via Eq.\,\eqref{eq:tau_lambda}.
For the acoustic modes at ${\bf q} = \Gamma = 0$, where $\omega ({\bf q \to 0}) \to 0$, the value for $\lambda_b (\Gamma)$ is obtained by averaging over values at the surrounding $\bf q$-points.
The scaling of lifetimes with $\omega_b^{-2} ({\bf q})$ used here is rooted in basic phonon theory as discussed in detail by Pomeranchuk and Herring, but it is not universal~\cite{Pomeranchuk.1941,Herring.1954}. Other scaling laws can be obtained by using different limiting assumptions that can, for example, depend on the crystal structure~\cite{Herring.1954}. However, the quadratic scaling used here is generally the strongest possible variation consistent with non-diverging thermal conductivities in the limit of dense Brillouin zone sampling irrespective of further limiting assumptions~\cite{Pomeranchuk.1941}.
Therefore, it leads to a firm upper bound for the effect of size extrapolation. 
Also due to the fact that this interpolation scheme is fully mode- and {\bf q}-resolved via the function $\lambda_b ( {\bf q})$ and incorporates the fully anharmonic lifetimes at commensurate ${\bf q}$-points, no systematic errors associated to the scaling and interpolation procedure have been observed in extended validation calculations covering several different crystal structures~\cite{Knoop.2022}.
 
\begin{figure}
	\includegraphics[width=\columnwidth]{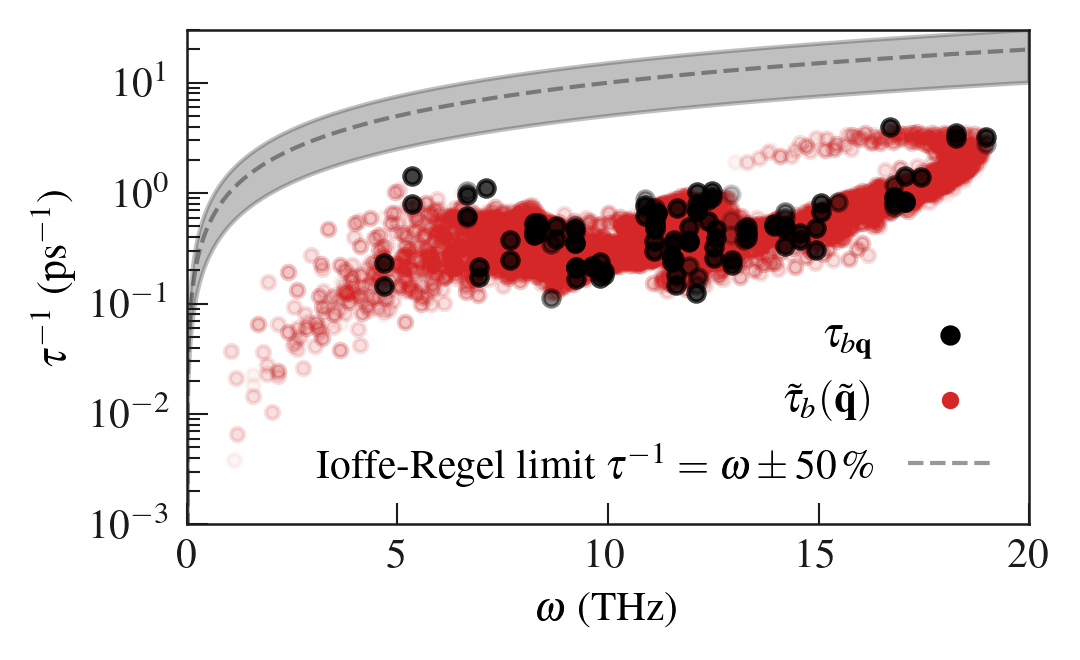}
	\caption{Scattering rates vs. frequency in \gls{mgo} at 300\,K after fitting Eq.\,\eqref{eq:G_b.approx} and symmetrizing using space-group operations as explained in the main text (black dots), and after interpolation to $20 \times 20 \times 20$ grid (red dots). Gray dashed: Ioffe-Regel limit $\tau^{-1} = \omega$ with 50\,\% margins.}
	\label{fig:tau_sq}
\end{figure}

For sampling the interpolated points $\set{ \tilde{\bf q} }$, we use even-numbered Monkhorst-Pack grids as implemented in phonopy~\cite{Monkhorst.1976,Togo.20154whg}, with a maximum grid size of $20 \times 20 \times 20$. The symmetrized lifetimes $\tau_\bq$ obtained from fitting Eq.\,\eqref{eq:G_b.approx}, as well as the interpolated lifetimes denoted by $\tilde{\tau}_b (\tilde{\bf q})$ are displayed as scattering rates ($\propto \tau^{-1}$) in Fig.\,\ref{fig:tau_sq}.
\begin{figure}
	\includegraphics[width=\columnwidth]{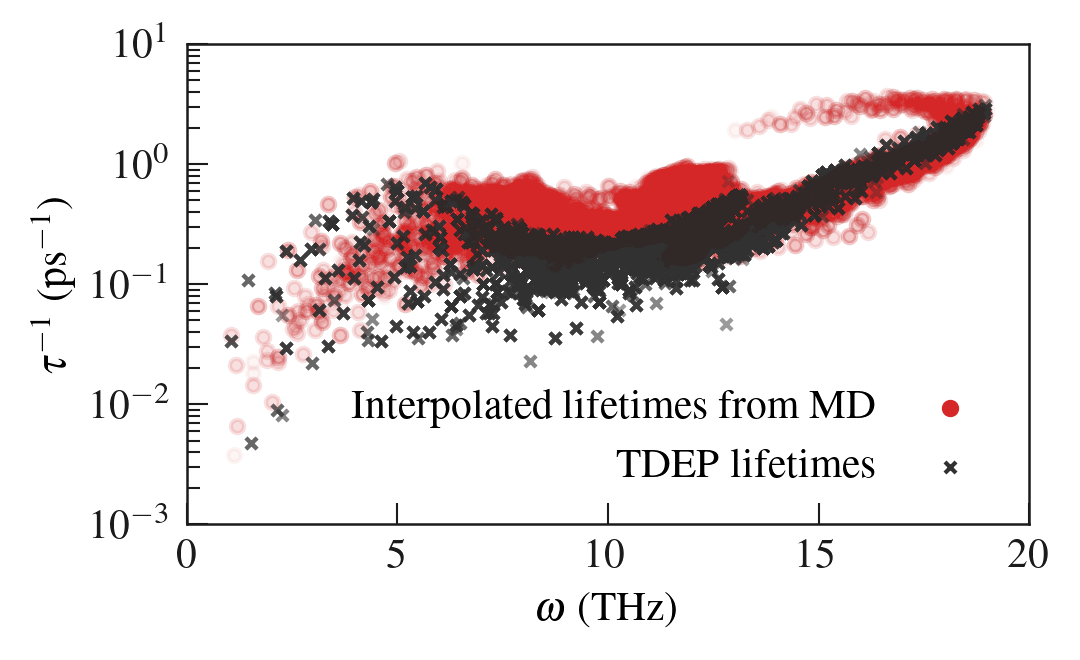}
	\caption{Scattering rates vs. frequency in \gls{mgo} at 300\,K. Comparisons of the lifetimes obtained via interpolation, cf.~Fig.~\ref{fig:tau_sq}, to those obtained using third-order perturbation theory with the TDEP code~\cite{Hellman.2011,Hellman.2013}.}
	\label{fig:tau_sq_TDEP}
\end{figure}
It is apparent that long-lived contributions stemming from modes with $\omega < 5$\,THz are introduced through the interpolation (red dots), effectively capturing modes with lifetimes $> 100$\,ps which is beyond the simulation time of 60\,ps. On the other hand, lifetimes for modes $\omega > 5$\,THz are already well-captured by the commensurate modes (black dots). 
This is further substantiated in Fig.~\ref{fig:tau_sq_TDEP}, which compares the lifetimes obtained by the described interpolation procedure with those obtained using third-order perturbation theory with TDEP force constants fitted to our MD simulations~\cite{Hellman.2013}. The difference observed between the perturbative and the interpolated scattering rates in the 7–12~THz range is attributed to higher-order anharmonic scattering, in line with the findings discussed for MgO at higher temperatures in Ref.~\cite{Puligheddu.2019} using a similar harmonic mapping procedure in larger supercells without interpolation. We also note that all scattering rates are well below the Ioffe-Regel limit $\tau^{-1} = \omega$~\cite{Sheng.1994}. This limit can be taken as a qualitative rule for estimating the validity of the phonon quasiparticle picture: Since the scattering rate is defined as the width of the phonon spectral function via Eq.\,\eqref{eq:G_b.approx}, $\tau_\bq^{-1} = 2 {\rm Im} \Sigma_\bq$, rates below this limit signify sharply peaked, 
well-defined quasiparticles~\cite{Simoncelli.2022,Caldarelli.2022bea}.

\subsection{Thermal conductivity extrapolation}
\begin{figure}
	\includegraphics[width=\columnwidth]{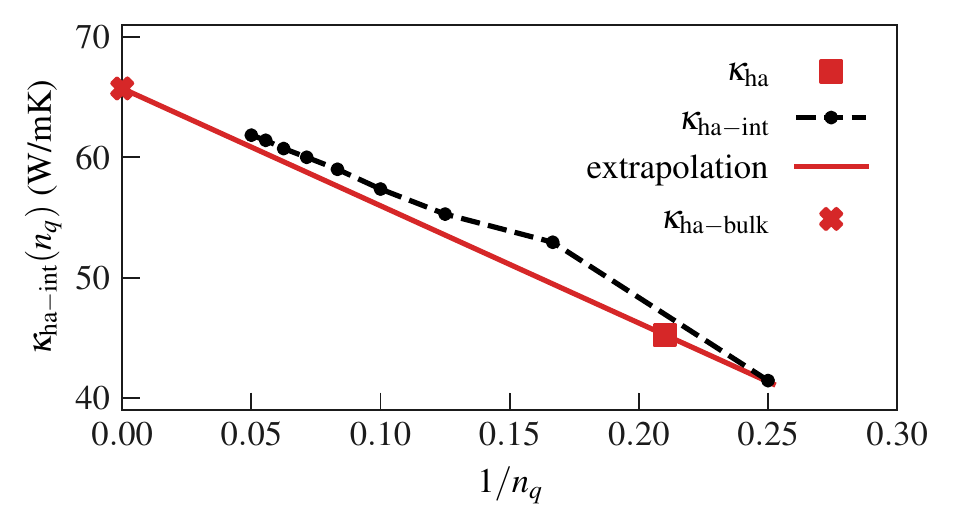}
	\caption{Size extrapolation correction to bulk limit computed from Eq.\,\eqref{eq:K.bte.correction} assuming linear convergence in $1 / n_q$, where $n_q$ is the number of {\bf q}-points per Cartesian direction. The offset between $\kappa_{\rm ha}$ and $\kappa_{\rm ha-int}$ arises because even grids are used for the extrapolation, whereas a $3 \times 3 \times 3$ supercell and respective grid of commensurate ${\bf q}$-points are used in the simulation.}
	\label{fig:imp.kappa.bte.correction}
\end{figure}

For the new, denser grid, an interpolated value,
\begin{align}
	\kappa_{\rm ha - int}^{\alpha \beta} (N_{\tilde{\bf q}}) = V k_{\rm B} \frac{N_{\bf q}}{N_{\tilde{\bf q}}} \sum_{b, \tilde{\bf q}} v^\alpha_b (\tilde{\bf q}) v^{\beta}_b (\tilde{\bf q}) \fD{\tilde{\tau}}_b (\tilde{\bf q})~,
	\label{eq:K.bte.correction}
\end{align}
can be obtained, where $N_{\tilde{\bf q}}$ is the number of points in the new grid, and the factor $N_{\bf q} / N_{\tilde{\bf q}}$ accounts for the increased number points. The bulk limit of Eq.\,\eqref{eq:K.bte.correction} is obtained by computing interpolated values for an increasing density of $\bf q$-points. 
The convergence of Eq.\,\eqref{eq:K.bte.correction} is approximately linear in $N_{\tilde{\bf q}}^{-1/3} \equiv 1 / n_q$, where $n_q$ is number of $\bf q$-points per Cartesian direction. The slope of this curve can therefore be used to extrapolate the value of $\kappa_{\rm ha}$ to bulk limit, as shown in Fig.\,\ref{fig:imp.kappa.bte.correction}.
With the extrapolated value $\kappa_{\rm ha-bulk}$, a correction can be obtained via
\begin{align}
	\delta \kappa_{\rm ha-correction} 
		= \kappa_{\rm ha-bulk} - \kappa_{\rm ha}~,
	\label{eq:K.correction}
\end{align}
from which the final result for the thermal conductivity is obtained via
\begin{align}
	\kappa^{\alpha \beta}_{\rm corrected}
		 = \kappa^{\alpha \beta} + \delta \kappa^{\alpha \beta}_{\rm ha-correction}~,
	\label{eq:K.corrected}
\end{align}
where $\kappa^{\alpha \beta}$ is the value from the \gls{aigk} simulation. The interpolation scheme effectively subtracts harmonic contributions to the thermal conductivity from vibrations commensurate with the supercell, and extrapolates them to the bulk limit, thereby including long-range contributions otherwise not present in the simulation cell. The size-corrected contributions are subsequently added back to the total thermal conductivity.

We note that several approximations are involved in the scheme outlined above, such as the assumption of exponential decay of the mode-energy autocorrelation function in Eq.\,\eqref{eq:G_b.approx}, or neglecting mode cross-correlations with $(b, {\bf q}) \neq (b', {\bf q}')$ in Eq.\,\eqref{eq:K.bte}. However, the dominant contribution to the size correction in Eq.\,\eqref{eq:K.correction} can be expected to come from low-frequency, long-lived phonons missing in the simulation cell, as shown in Fig.\,\ref{fig:tau_sq}, for which the approximations listed above are well justified. The shorter-range contributions of modes that interact more strongly are fully captured on the \emph{ab initio} level, for which the lifetimes are only weakly frequency-dependent, as seen in Fig.\,\ref{fig:tau_sq} when focusing on the regime where $\omega > 5\,{\rm THz}$.

We also note in passing that the force constants used for size extrapolation could also be obtained via finite differences as,~e.\,g.,~implemented in phonopy~\cite{Parlinski.1997,Togo.20154whg,Carreras.2017}. However, especially for quite harmonic materials such as \gls{mgo}, we didn't find the extrapolation scheme to be sensitive to subtle differences in the force constants used to describe the auxiliary (effective) harmonic model. 
Furthermore, effective harmonic models extend the applicability of phonon theory to dynamically unstable high-temperature phases, for example in 
SrTiO$_3$~\cite{Tadano.2015}, PdH~\cite{Paulatto.2015}, or ZrO$_2$~\cite{Carbogno.2014,Carbogno.2017}. A detailed account of the feasibility  of the size-extrapolation scheme  presented here for these systems is, however, beyond the scope of the current work.

\section{Results for magnesium oxide \label{sec:time_convergence}}
After we have seen how the cutoff time $\tcut$ in Eq.\,\eqref{eq:kappa.trunc.t} can be obtained, and finite-size errors can be corrected, we discuss the convergence of presented scheme as a function of the simulation time $t_0$ in Eq.\,\eqref{eq:hfacf}. We do this for the case of \gls{mgo} for three independent trajectories of 60\,ps length each. We truncate every trajectory in 10\,\% steps down to a length of 6\,ps, and apply the workflow presented in the previous sections to each of the truncated trajectories. 
\begin{figure}
	\includegraphics[width=\columnwidth]{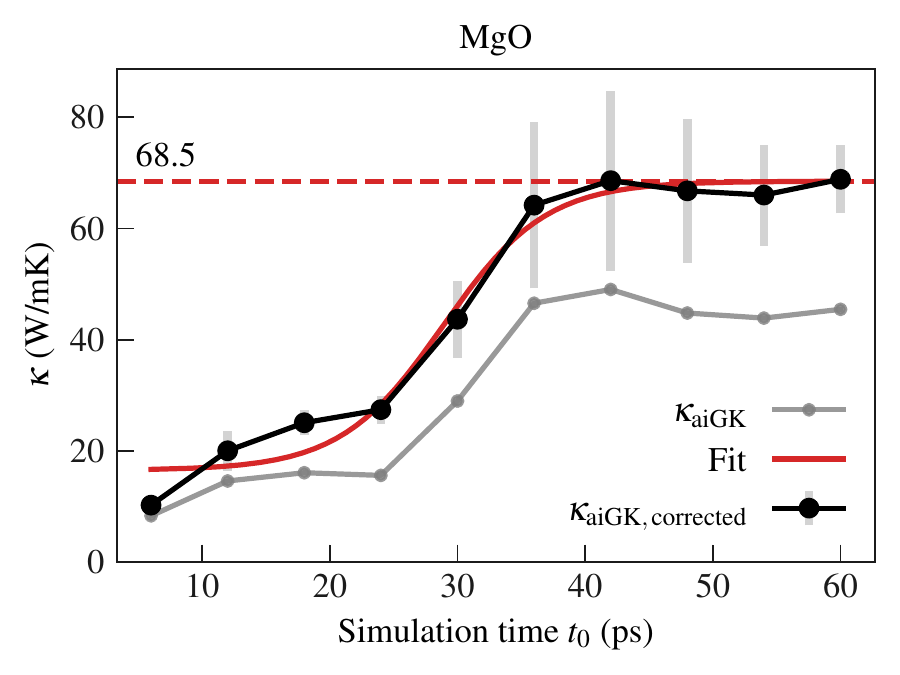}
	\caption{Thermal conductivity $\kappa$ as function of the simulation time $t_0$ as defined in Eq.\,\ref{eq:teff}. Values are given as the ensemble average over three independent trajectories. The error bars are computed according to Eq.\,\eqref{eq:kappa.error} as the standard error of the ensemble average. The blue curve is a logistic curve defined in Eq.\,\eqref{eq:f_logistic} fitted to the $\kappa$ values, the dashed blue curve is the infinite time limit of the fitted function. Gray dots represent the thermal conductivity as given by the simulation without the size-correction scheme.
	Please note that the values for $\kappa$ shown here cannot be directly compared to the value displayed in Fig.\,\ref{fig:imp.hfacf.kappa.1} or \ref{fig:imp.kappa.bte.correction}, because the latter only show single components of single runs, which can vary substantially from the total average.}
	\label{fig:kappa_convergence.mgo}
\end{figure}
Figure~\ref{fig:kappa_convergence.mgo} shows that the thermal conductivity converges to a plateau after about 40\,ps, where the value of $\kappa$ stays constant within the error bars. The overall shape of the curve can be described as follows: Simulations shorter than 20\,ps sample the early decay of the \gls{hfacf} which contribute about 30\,W/mK to the total thermal conductivity After a simulation time of 25\,ps, the late decay of the \gls{hfacf} is sampled, contributing more than double the amount to the total thermal conductivity of $68.8 \pm 6.1$\,W/mK after the total simulation time. In the plot, this two-step behavior is approximated by a logistic function 
\begin{align}
	f(t) 
		= \frac{L}{1 + \exp \left(-\frac{(t-t_{\rm inflection})}{\tau} \right)} + f_0~,
	\label{eq:f_logistic}
\end{align}
which captures the second super-linear increase in $\kappa$ at $t_{\rm inflection} \simeq 29\,{\rm ps}$, and models the correct asymptotic behavior for long times. The asymptotic value of $\kappa (t_0 \to \infty) = 68.5$\,W/mK agrees very well with the value after 60\,ps of $\kappa (60\,{\rm ps}) = 68.8 \pm 6.1$\,W/mK. As the largest lifetime in the simulation corresponds to $\tau \approx 15.5$\,ps, as highlighted in Fig.\,\ref{fig:G_s}, we do not expect a significant increase of $\kappa (t_0)$ after this simulation time. We therefore conclude that the simulation time of $t_0 = 60$\,ps can be considered converged, and that Eq.\,\eqref{eq:f_logistic} can be used to model the late increase of $\kappa (t_0)$, in line with the division into short and long processes commonly discussed in the literature~\cite{Ladd.1986,Kaburaki.1998,McGaughey.2004lnu}. Furthermore, we note that the size extrapolation increases the value from $\kappa = 45.5$\,W/mK to $\kappa_{\rm corrected} = 68.8$\,W/mK,~i.\,e.,~the value increases significantly by about 50\,\%.

We like to point out that we did not discard an initial time from the {\it NVE} simulations to allow for further equilibrating after the thermostat is switched off, as is common practice in the literature on \gls{gk} simulations~\cite{McGaughey.2004lnu}. We did not find this procedure to be necessary: The truncation of simulation times displayed in Fig.\,\ref{fig:kappa_convergence.mgo} was performed such that the \emph{early} time steps in the simulation were discarded,~i.\,e., a simulation time of $t_0 = 54$\,ps corresponds to discarding the first 6\,ps from the trajectory. Since discarding 6\,ps or more did not change the result significantly, its effect can be assumed to be minor for \gls{aigk} simulations.

\subsection{Comparison to literature \label{sec:literature.mgo}}
We conclude the discussion for \gls{mgo} by comparing to available experimental and theoretical references. These references are listed in Tab.\,\ref{tab:references.MgO}. 
\begin{table}[ht]
  \centering
  \begin{tabular}{lc}
    \toprule
    Reference & Thermal conductivity \\
    & at 300\,K (W/mK) \\
    \midrule
  Experiment~\cite{Slack.1962,Touloukian.1971,MacPherson.1983,Andersson.1986,Katsura.1997,Dalton.2013,Hofmeister.2014} &  50-65 \\
    de Koker 2010 (LDA)~\cite{Koker.2010} & $\approx 75^{\,\dagger}$ \\
    Stackhouse et al. 2010 (LDA)~\cite{Stackhouse.2010} & $58 \pm 6^{\,\dagger}$ \\
    Tse et al. 2018 (PBE)~\cite{Tse.2018} & $70.3 \pm 8.9$ \\
    Dekura and Tsuchiya 2017 (LDA)~\cite{Dekura.2017} & $\approx 54$ \\
    Plata et al.~2017 (PBE)~\cite{Plata.2017} & $54.06$ \\
    Xia et al.~2020 (PBE)~\cite{Xia.2020} & $50.1-58.7$ \\
    This work & $68.8 \pm 6.1$ \\
    \bottomrule
    \vspace{.5em}
  \end{tabular}
  \caption{Reference values for the thermal conductivity of periclase \gls{mgo} at ambient conditions. Values marked with $\dagger$ are extrapolated values using data from higher temperatures using Eq.\,(17) in Ref.\,\cite{Koker.2010} and Eq.\,(5) in Ref.\,\cite{Stackhouse.2010}, respectively. See also discussion in Ref.\,\cite{Haigis.2012}. For the \emph{ab initio} studies, the level of theory is indicated in parentheses: local-density approximation (LDA)~\cite{Ceperley.1980}, or the \gls{gga} parametrized by Perdew, Burke, and Ernzerhof~(PBE)~\cite{Perdew.1996}.
  }
  \label{tab:references.MgO}
\end{table}
The given references show a significant spread. On the experimental side, the main source of uncertainty arises from different sample quality, different measurement techniques, and the fact that thermal conductivity is usually not the direct observable, but inferred from thermal diffusivity measurements and other material parameters which may imply additional sources of error~\cite{Hofmeister.2014}. The agreement between our aiGK simulation and experiment is satisfactory. While our thermal conductivity is larger than that of the listed experiments, this is to be expected since isotope effects are not included in our simulations. The listed measurements on the other hand are not performed for isotopically pure \gls{mgo}.
Neglecting these isotope scattering effects can lead to increases of thermal conductivity. Using TDEP with force constants fitted to our PBEsol trajectories~\cite{Hellman.2011,Hellman.2013}, we obtain an increase of 35\% in the single-mode relaxation time approximation and of 46\% when solving the full linearized \gls{bte}, in line with the 46\,\% increase reported in Ref.~\cite{Tang.2010} using the LDA functional. Corrected by this factor, our result would be $47.1 \pm 4.2$\,W/mK, only slightly below and within error of the most recent experimental results of 50.1\,W/mK obtained by Hofmeister~\cite{Hofmeister.2014}.

On the theoretical side, we compare to three other approaches based on \gls{aimd} simulations by de~Koker~\cite{Koker.2009,Koker.2010}, Stackhouse, Stixrude, and Karki~\cite{Stackhouse.2010}, and Tse et al.\cite{Tse.2018}. The quantitative agreement with de~Koker and Stackhouse et al. is acceptable, given that different xc~functionals and otherwise computational settings were used. We note that the higher values found by de Koker are a little surprising, given that smaller supercells were used, and no size extrapolation. However, the value of $\kappa \approx 75$\,W/mK listed in Tab.\,\ref{tab:references.MgO} is an extrapolation from higher temperatures, where finite-size effects are likely less important. The agreement with the study by Tse et al.~\cite{Tse.2018} based on the Einstein relation introduced in Ref.\,\cite{Kinaci.2012} is very good.

The other theoretical works are based on perturbative \gls{bte} approaches~\cite{Broido.2007}. The listed references use three-phonon scattering to compute phonon lifetimes, the lowest value reported by Xia et al.~\cite{Xia.2020} is obtained by additionally including fourth-order scattering which further reduces lifetimes in \gls{mgo}~\cite{Feng.2017}. The \gls{bte} approaches based on third-order scattering listed here account for isotope scattering and are therefore consistently lower than our \gls{aigk} value, since isotope scattering is more pronounced than higher-order phonon-phonon scattering in \gls{mgo}.

Given the comparatively large uncertainty inherent to thermal conductivity measurement and simulation, the agreement between aiGK and the available literature can therefore be considered satisfactory. The discussion for \gls{mgo} further shows that \gls{aigk} can be used for mostly harmonic materials with considerable phonon lifetimes \emph{when} a suitable extrapolation scheme is employed.

\section{Results for copper iodide \label{sec:cui}}
Next, we apply the scheme as presented above to marshite \gls{cui}, a strongly anharmonic material which becomes a superionic conductor above 643\,K~\cite{Boyce.1980,Boyce.1981}. 

\begin{figure}
	\includegraphics[width=\columnwidth]{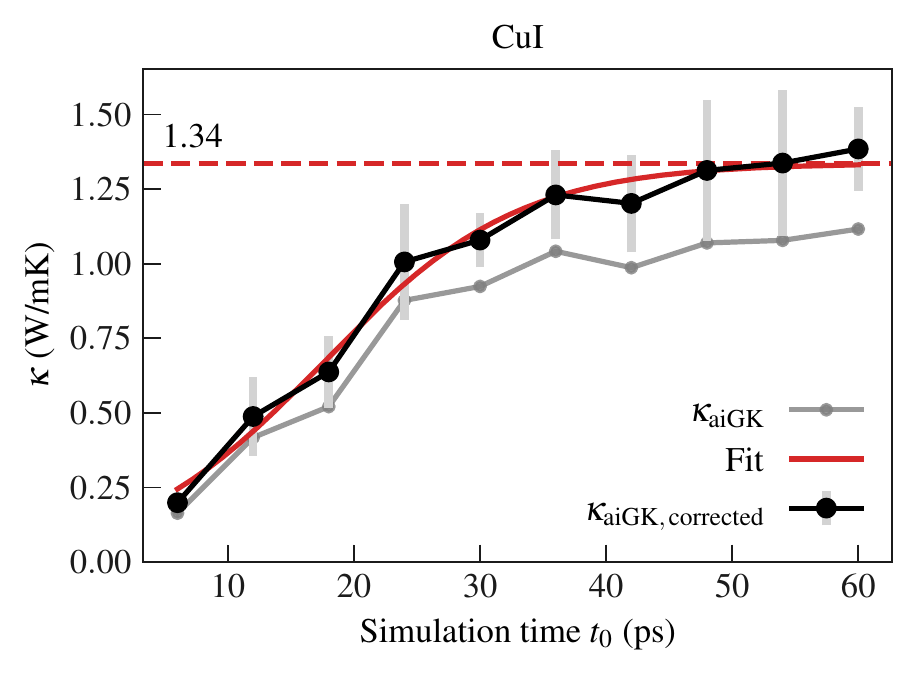}
	\caption{Thermal conductivity $\kappa$ for \gls{cui} as function of the simulation time $t_0$. Curves and symbols are equivalent to those defined in Eq.\,\ref{fig:kappa_convergence.mgo}.}
	\label{fig:kappa.convergence.CuI}
\end{figure}

\begin{figure}
	\includegraphics[width=\columnwidth]{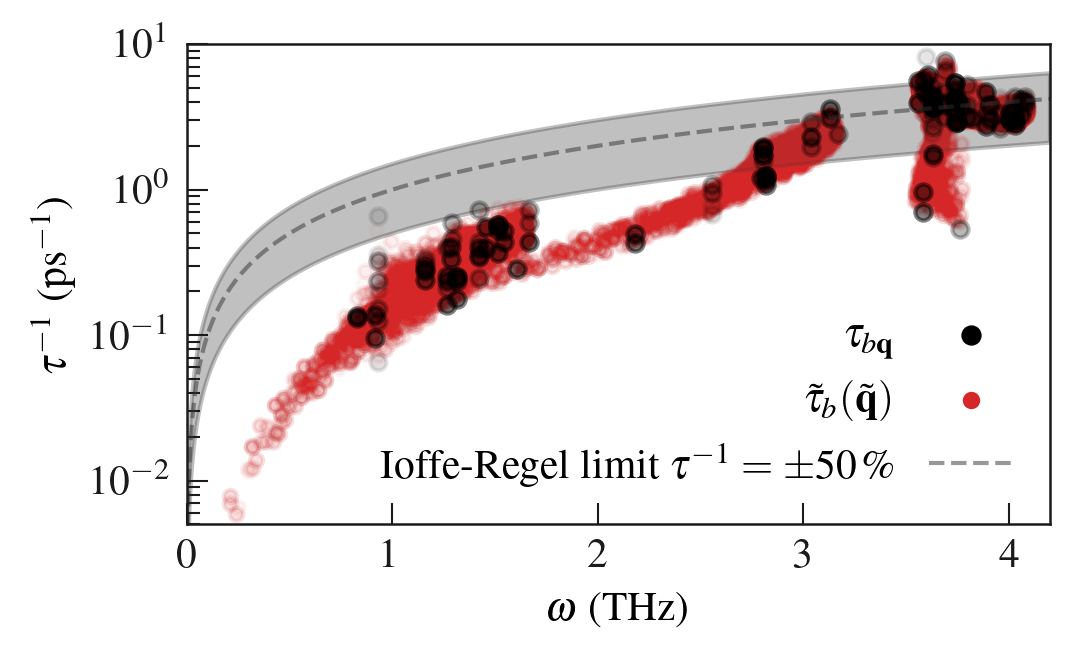}
	\caption{Scattering rates vs. frequency in \gls{cui} at 300\,K after fitting Eq.\,\eqref{eq:G_b.approx} and symmetrizing using space-group operations as explained in the main text (black dots), and after interpolation to $20 \times 20 \times 20$ grid (red dots). Gray dashed: Ioffe-Regel limit $\tau^{-1} = \omega$ with 50\,\% margins.}
	\label{fig:tau_sq.cui}
\end{figure}

The final ensemble-averaged thermal conductivity as function of the simulation time is displayed in Fig.\,\ref{fig:kappa.convergence.CuI}.
Fitting the logistic function in Eq.\,\eqref{eq:f_logistic} to $\kappa (t)$ as before, we can pinpoint the second superlinear increase in $\kappa (t)$ to $t_{\rm inflection} \simeq 18\,{\rm ps}$, although this increase is visually less pronounced than in \gls{mgo}. The early increase is dominated by the considerably stronger scattering and therefore shorter lifetimes of modes in the optical range, as shown in Fig.\,\ref{fig:tau_sq.cui}.
As the lifetimes in the lower frequency range are comparable to those of \gls{mgo}, we conclude that the simulation time of 60\,ps is sufficient for \gls{cui} as well. The size extrapolation scheme increases the thermal conductivity from $\kappa = 1.12$\,W/mK to $\kappa_{\rm corrected} = 1.38$\,W/mK,~i.\,e.,~by about 23\,\%, which is in line with the assumption that finite-size effects become less important in strongly anharmonic materials.

As \gls{cui} is less abundant than \gls{mgo}, it is less frequently covered in the literature. The available reference is summarized in Tab.\,\ref{tab:exp.CuI}.
\begin{table}[ht]
  \centering
  \begin{tabular}{lc}
    \toprule
    Reference & Thermal conductivity \\
    & at 300\,K (W/mK) \\
    \midrule
    CRC Handbook~\cite{Perry.2016} (experiment, bulk) & 1.68 \\
    Yang et al.~\cite{Yang.2017} (experiment, thin film) &  0.55 \\
    Togo et al.~\cite{Togo.2015} (theory) & 6.55--7.22 \\
    This work & $1.38 \pm 0.14$ \\
    \bottomrule
    \vspace{.5em}
  \end{tabular}
  \caption{Experimental values and one theoretical reference for the thermal conductivity of marshite \gls{cui} at ambient conditions. The value from Yang et al. marked by $\dagger$ is from a thin film experiment, and therefore can be regarded as a lower bound of the bulk thermal conductivity~\cite{Yang.2017}.}
  \label{tab:exp.CuI}
\end{table}
We slightly underestimate the CRC Handbook reference of $1.68$\,W/mK~\cite{Perry.2016}, but clearly above the thin-film reference of about $0.55$\,W/mK as reported by Yang and coworkers~\cite{Yang.2017}, which we take as a firm lower limit to the intrinsic thermal conductivity of \gls{cui} due to boundary scattering. It is noteworthy that an earlier computational investigation based on Boltzmann transport theory by Togo and coworkers reaches a much higher value of thermal conductivity in \gls{cui} of about 7\,W/mK~\cite{Togo.2015}. We propose the following explanation based on the findings for other strongly anharmonic zincblende compounds presented by Xia and coworkers in Ref.\,\cite{Xia.2020}: The authors showed that higher-order phonon scattering can limit the thermal conductivity in zincblende compounds considerably, and including only third-order scattering can overestimate $\kappa$ by up to 450\,\% in the case of HgTe, a compound which is less anharmonic than \gls{cui} according to the quantification scheme presented in Ref.\,\cite{Knoop.2020}. Since the aiGK method is non-perturbative, anharmonic scattering up to arbitrary order is naturally included, which explains the variance with BTE results using third-order scattering only. This is further supported by comparing the scattering rates displayed in Fig.\,\ref{fig:tau_sq.cui} to the Ioffe-Regel introduced earlier~\cite{Sheng.1994}: Nearly all modes in the optical part of the spectrum $>2.5$\,THz approach or exceed the range $\tau^{-1} = \omega$. Several modes in the range 1-2\,THz approach this limit. As discussed in Sec.\,\ref{sec:extrapolation.interpolation} for MgO, a scattering rate in this range signifies a strong broadening of the phonon spectral function  beyond the perturbative regime which is 
defined for $\tau^{-1} \ll \omega$~\cite{Simoncelli.2022,Caldarelli.2022bea}.

\section{Remark on simulation times \label{sec:simulation.times}}
\begin{figure}
	\includegraphics[width=\columnwidth]{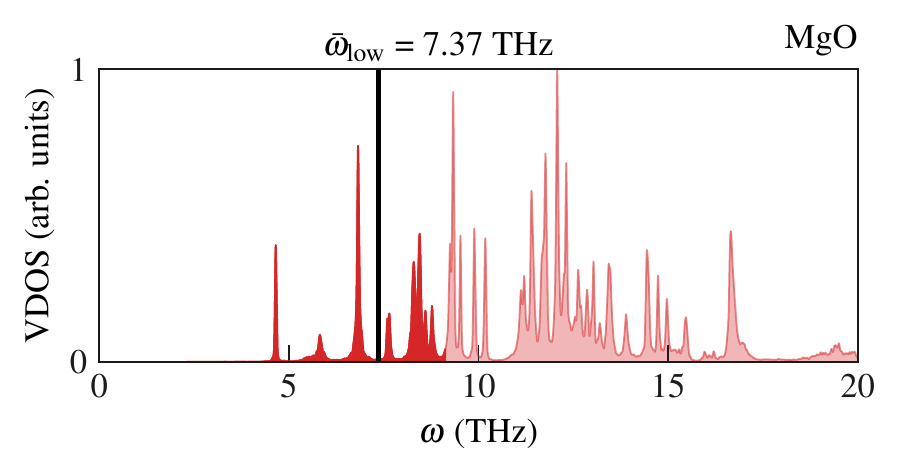}
	\includegraphics[width=\columnwidth]{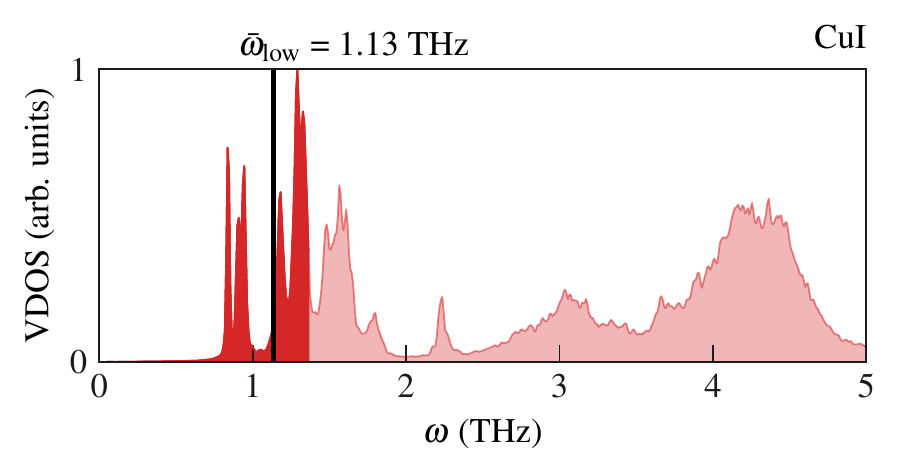}
	\caption{Significant portion of the vibrational density of states (VDOS)~\cite{Dove.1993}, and representative low frequency $\wmin$ for \gls{mgo} (upper panel) and \gls{cui} (lower panel). Note the different limits on the x-axis.
	}
	\label{fig:vdos}
\end{figure}

It is clear that the minimal necessary simulation time $t_0$ is material-dependent and needs to be checked in each study. It may therefore come as a surprise that 60\,ps turned out to sufficient both for \gls{mgo} and \gls{cui}, although their dynamical and anharmonic properties are quite different, and one might \emph{a priori} expect much longer simulation times to be necessary for the more harmonic \gls{mgo}. However, when inspecting the vibrational properties of both materials, one can infer that the \emph{effective} simulation time for \gls{mgo} is indeed much longer than for \gls{cui}: We define the dimensionless effective simulation length via
\begin{align}
	\teff = t_0 \cdot \wmin~,
	\label{eq:teff}
\end{align}
where $t_0$ is the simulation time, and $\wmin$ is a characteristic frequency for the slow degrees of freedom of the system, motivated by the fact that heat transport is usually dominated by these slow processes. We choose $\wmin$ as the mean frequency of the lowest 20\,\% of the vibrational spectrum as shown in Fig.\,\ref{fig:vdos}, but emphasize that the argument is not sensitive to this somewhat arbitrary choice. 
Using the effective simulation length as defined in Eq.\,\eqref{eq:teff}, we see that the simulation for \gls{mgo} ($\teff = 442.2$) is effectively 6.5 times longer than that for \gls{cui} ($\teff = 67.8$), due to the stiffer bonding and therefore faster vibrations present in \gls{mgo}. Noting that lifetimes tend to decrease with increasing frequency as argued earlier~\cite{Herring.1954}, this explains why the longest lifetimes in the simulation cell are of the same order of magnitude in \gls{mgo} and \gls{cui} as shown in Fig.\,\ref{fig:tau_sq} and~\ref{fig:tau_sq.cui}, despite the much stronger anharmonic character of \gls{cui}.

For novel materials, we therefore suggest to estimate the necessary simulation times based on the vibrational spectrum and lifetime estimates according to Eq.\,\eqref{eq:G_b.approx}. Special care must be taken for slow but harmonic materials, for which the necessary simulation times might easily be much longer than those reported here, consistent with GK studies based on empirical force fields for harmonic materials like silicon~\cite{Volz.1999}.

\section{Conclusion and outlook \label{conclusion}}
We have presented and applied an implementation of \gls{aigk} simulations based on the first-principles heat flux introduced in Ref.\,\cite{Carbogno.2017}. Systematically removing noise from the \gls{hfacf} allows to estimate cutoff times in a numerically robust way. We also presented a detailed account of our updated size-extrapolation scheme based on mapping the \gls{aimd} trajectories to an (effective) harmonic model. The scheme was applied to \gls{mgo} and \gls{cui}, two simple binary systems of quite different harmonic character: \gls{mgo} is an example for a stiffly-bonded, harmonic material with quite high thermal conductivity for a rock salt compound, whereas \gls{cui}  is a strongly anharmonic compound that dynamically destabilizes and becomes superionic conducting at higher temperatures~\cite{Knoop.2020,Boyce.1980}. Good agreement with the available literature is found for \gls{cui}, and for \gls{mgo} when correcting for isotope effects.

The presented scheme and its implementation in FHI-vibes makes performing, post-processing, and analyzing \gls{aigk} simulations much more straightforward than previously possible. The reduced human intervention enables to study heat transport in materials across materials space, in particular for strongly anharmonic or complex materials, as discussed in Ref.\,\cite{Knoop.2022}.
Furthermore, approaches such as those based on \gls{bte} can be systematically benchmarked against non-perturbative results obtained from \gls{aigk} in the future.

We note in passing that the presented approach can equally be applied to GK studies based on empirical or machine-learned force fields~\cite{Sosso.2012,Korotaev.2019,Qian.201965,Li.2020,Fan.2021,Verdi.2021,Langer.2022a}.
Also in these cases, the presented strategies that rely on physically motivated approaches and parameters can be helpful to obtain more stable and systematic data from potentially noisy simulations.

\section*{Data and code availability}
The presented method is implemented and available in the open source package FHI-vibes~\cite{Knoop.2020cx}. The package builds on the atomic simulation environment (ASE)~\cite{Larsen.2017}. The data and scripts used to create the plots are made available via figshare~\cite{figshare}.
The raw DFT calculations including input and output files are accessible via NOMAD. \cite{nomad1}.

\section*{Acknowledgments}
This project was supported by the NOMAD Center of Excellence (European Union’s Horizon 2020 research and innovation program, grant agreement No. 951786), the ERC Advanced Grant TEC1p (European Research Council, grant agreement No. 740233), and the North-German Supercomputing Alliance (HLRN). F.\,K. acknowledges support from the Swedish Research Council (VR) program 2020-04630, and the Swedish e-Science Research Centre (SeRC).
F.\,K. would like to thank Stefano Baroni and Federico Grasselli for inspiration and fruitful discussions related to \gls{gk} theory, and Marcel Langer for useful feedback on the method and manuscript.

\bibliographystyle{apsrev4-1}

\end{document}